# Conditional Pulse Nulling Receiver for Multi-pulse PPM and Binary Quantum Coding Signals

Yuan Zuo[1, 2], Tian Chen[1, 2], and Bing Zhu[1, 2]

[1] Department of Electronic Engineering and Information Science, University of Science and Technology of China, Hefei 230027, China

[2] Key Laboratory of Electromagnetic Space Information, Chinese Academy of Sciences, Hefei 230027, China

Email: zuoyuan@mail.ustc.edu.cn; chentian@mail.ustc.edu.cn; zbing@ustc.edu.cn

*Abstract* —Conditional pulse nulling (CPN) receiver is proposed by Dolinar to discriminate pulse position modulation (PPM) signals. The receiver using a beam splitter and an on-off photon detector can outperform the standard quantum limit (SQL) for PPM signals. In this paper, we apply this receiver to multi-pulse PPM (MPPM) and binary quantum coding signals, and use dynamic programming algorithm to optimize the control strategy. The MPPM signals is used to improve the band-utilization efficiency, and the binary quantum coding make the communication able to correct the error. Numerical simulation results show that the CPN receiver with optimized strategy can also outperform SQL for both MPPM and binary quantum coding signals.

## I. Introduction

In an optical communication protocol, information are encoded into different coherent states. The performance of discriminating different coherent states is limited by the fact that coherent states are nonorthogonal instinctively. Classical measurement method such as homodyne detection, heterodyne detection and direct detection are limited by this instinct also called shot noise. These limits are known as standard quantum limit (SQL). However quantum measurement sets more fundamental limits known as Helstrom limit [1]. For binary modulation, e.g on-off keying (OOK) and binary phase shift keying (BPSK), Dolinar receiver achieves the Helstrom limit theoretically [2]. And for phase shift keying (PSK) modulation, some structures are presented and their performances are lower than SQL but not achieve Helstrom limit [3-6].

Pulse position modulation (PPM) signals encode the information to the position of a pulse in the symbol time. This modulation is widely used in free space optical communications like satellite-to-satellite and satellite-to-earth links. For this signals, Dolinar proposed a conditional pulse nulling (CPN) receiver to reduce the average error probability below the SQL [7]. A nature extension of single-pulse PPM is the use two or more pulse in each symbol time, which is called multi-pulse PPM (MPPM). The MPPM signals enable carrying more information in one symbol, and increasing the band-utilization efficiency [8]. In classical MPPM system, direct detection is applied to detect these signals, which has been studied by M. K. Simon [9].

Quantum channel code make it possible to transmit signals through imperfect channel with error-correcting ability. Moreover, quantum channel code enable both superadditivity and achieving Holevo capacity [10-12]. Square root measurement, sequential measurement and structured joint-detection receiver are considered to be able to achieve Holevo capacity [10, 13, 14]. But it is difficult to implement those receivers with realistic optical devices.

Conditional pulse nulling receiver is used to discriminate PPM signals by Dolinar. It nulls the PPM signals depending on the previous measurement. This receiver outperforms the SQL and approach the theoretical Helstrom limit. This result has been demonstrated experimentally [7, 15]. In this paper, we improve this structure to discriminate MPPM and binary quantum coding signals. The key issue is how to determine the best control strategy in signal time duration. To solve this question, we use a dynamic programming algorithm which has been used for optimizing the control strategy for PPM conditional pulse nulling receiver to further reduce the error rate [16]. Numerical results show that this receiver can outperform the SQL given by classical detection method for both MPPM and quantum code signals. Besides, it can be implement by only linear optical devices, such as a beam splitter and an on-off photon detector.

## II. Signals And Receiver

### A. Multi-pulse PPM Signals

In quantum theory, coherent light pulse is represented as coherent states $|\alpha\rangle$ in a Hilbert space $\mathcal{H}$. Its average photon number is $|\alpha|^2$. For Multi-pulse PPM (MPPM) signals, there are two or more pulse in a symbol duration. It is represented as product state in the tensor Hilbert space $\mathcal{H}^{\otimes M}$, e.g. $|\alpha\rangle|0\rangle|0\rangle|\alpha\rangle$. The state $|0\rangle$ is vacuum state, means no pulse in this time slot. Generally MPPM signals with $M$ time slots can be written as

$$|\gamma\rangle = |\gamma_1\rangle|\gamma_2\rangle\cdots|\gamma_M\rangle, \quad \gamma_i = \begin{cases} \alpha \\ 0 \end{cases}. \quad (1)$$

If there are $L$ pulses in all $M$ time slots, we called them $L$-pulse $M$-ary PPM (abbreviated as $L$-$M$-PPM) signals. It is easy to calculate that there are $\binom{M}{L} = \frac{M(M-1)\cdots(M-L+1)}{L!}$ different $L$-$M$-PPM signals in total. For the simple case $M = 4$ and $L = 2$, the number of 2-4-PPM signals is 6. Those signals are 1100, 1010, 1001, 0110, 0101, and 0011, where we use 0 for $|0\rangle$ and 1 for $|\alpha\rangle$.

When information are sent through optical communication channel, the transmitter encodes message into a sequence of symbols. Then the transmitter maps the symbols to the MPPM signals. The signals pass through the channel, which is ideal in our assumption, then arrive the receiver. The receiver measure the signals by a set of positive operator value measurement (POVM) and extract the original message. In traditional MPPM signals optical communication system, the direct detection (DD) is applied and decode by a maximum-likelihood criterion [9].

### B. Binary Quantum Coding Signals

Quantum channel coding theorem [10, 11] shows that the ultimate capacity limit for pure state system is bounded by Holevo capacity

$$C = \max_{p_i}\left[ H\left(\sum_i p_i |\psi_i\rangle\langle\psi_i|\right)\right]. \quad (2)$$

Where $H(\rho) = -\mathrm{Tr}\,\rho \log_2 \rho$ is the von Neumann entropy for the system with density matrix $\rho$, and $p_i$ is prior probability of the state $|\psi_i\rangle$. Besides, this limit can be achieved by joint detection over long code-word blocks. Some code-word such as random code and polar code have been investigated to achieve this limit. And square root measurement is used for those signals detection [10, 17].

In this paper, we only focus on binary quantum coding signals

$$|\gamma\rangle = |\gamma_1\rangle|\gamma_2\rangle\cdots|\gamma_M\rangle, \quad \gamma_i = \begin{cases} 0 \text{ or } \alpha, & \text{for OOK} \\ -\alpha \text{ or } \alpha, & \text{for BPSK} \end{cases}. \quad (3)$$

Where $M$ is the code length. The OOK modulation quantum coding signals are similar to MPPM signal. They have both several pulses in $M$ time slots. The difference is the pulse number is variable in OOK quantum coding signals. In traditional optical communication system, receiver direct detects each time slot for OOK modulation and uses homodyne detection for BPSK modulation. After outputting all time slots result of a code-word, the receiver uses a maximum-likelihood criterion to determine the most likely code [9, 18].

### C. Conditional Pulse Nulling Receiver

Conditional pulse nulling receiver structure is shown in Fig. 1 ($a$). In each time slot, signal is displaced by a local optical field using a beam splitter, which is described by a displacement operator $D(\beta)$. Then the displaced optical field is detected by an on-off detector. The result of this time slot feedbacks to change local optical field in next time slot. If detector received photons in this time slot, then change the local optical field to set the displacement operator parameter as $\beta^0$ in the next slot. Otherwise set the parameter as $\beta^1$. In generally, the feedback strategy can be described using a decision tree (see Fig. 1 ($b$)). In the case of 4-PPM $\{|\alpha\rangle|0\rangle|0\rangle|0\rangle$, $|0\rangle|\alpha\rangle|0\rangle|0\rangle$, $|0\rangle|0\rangle|\alpha\rangle|0\rangle$, $|0\rangle|0\rangle|0\rangle|\alpha\rangle\}$, $\{\beta, \beta^1, \text{ and } \beta^{11}\}$ are all set to $-\alpha$, which means nulling pulse in these slots. Ideally some leaves of the tree are unreachable. For the unideal condition, receiver can use maximum posterior probability (MAP) decision rule [7].

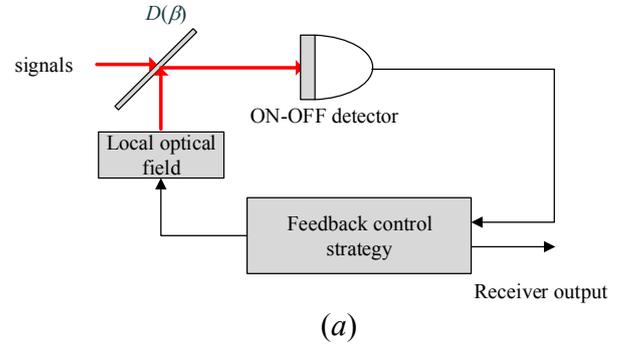

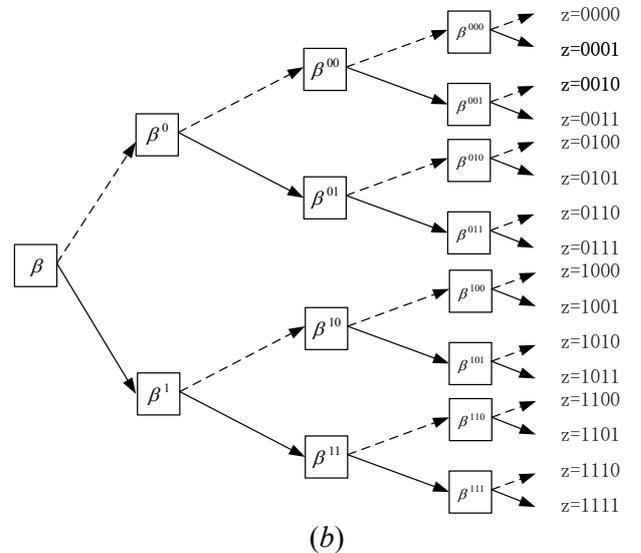

Fig. 1. ($a$) Condition pulse nulling receiver structure schematic diagram. ($b$) Strategy tree of conditional nulling receiver for the signals with 4 time slots, such as 4-PPM, 2-4-PPM and binary quantum coding signals. Dash arrow means no photon click while solid arrow means photon click.

## III. STRATEGY AND NUMERICAL RESULTS

### A. Receiver Strategy

Both classical receivers and our receiver measure the signals slot by slot. In each slot, the classical receivers for MPPM and OOK quantum coding signals direct detect whether any photons present using a photon detector. An ideal direct detection can be expression a pair of POVM operators

$$\Pi_0 = |0\rangle\langle 0|, \ \Pi_1 = I - |0\rangle\langle 0|. \quad (4)$$

The conditional probabilities for states are

$$\begin{aligned} p_{0|0} &= \text{Tr}\left[\Pi_0 |0\rangle\langle 0|\right] = 1, \\ p_{1|1} &= \text{Tr}\left[\Pi_1 |\alpha\rangle\langle \alpha|\right] = 1 - e^{-|\alpha|^2}. \end{aligned} \quad (5)$$

In our conditional pulse nulling receiver, for MPPM and OOK quantum coding signals we choose detection method from nulling operator and direct detection. When the nulling operation is performed, a displacement $D(-\alpha)$ is applied to the signal state. In this case, the vacuum state $|0\rangle$ is displaced to $|-\alpha\rangle$ and the coherent state $|\alpha\rangle$ is displaced to $|0\rangle$. The conditional probabilities using nulling operator are

$$\begin{aligned} p_{0|0} &= \text{Tr}\left[\Pi_0 D(-\alpha)|0\rangle\langle 0|D(-\alpha)^\dagger\right] = e^{-|\alpha|^2}, \\ p_{1|1} &= \text{Tr}\left[\Pi_1 D(-\alpha)|\alpha\rangle\langle \alpha|D(-\alpha)^\dagger\right] = 0 \end{aligned} \quad (6)$$

For BPSK quantum coding signals, we choose which state to be nulled. When nulling state $|-\alpha\rangle$, the conditional probabilities are

$$p_{0|0} = 1, \ p_{1|1} = 1 - e^{-4|\alpha|^2}. \quad (7)$$

Where we denote 0 for $|-\alpha\rangle$ and 1 for $|\alpha\rangle$. While nulling state $|\alpha\rangle$, the conditional probabilities change in

$$p_{0|0} = 1 - e^{-4|\alpha|^2}, \ p_{1|1} = 0. \quad (8)$$

After each slot, the on-off photon detector output 1 for receiving photons and 0 for no photon. We use $z_k$ to denote the $k$-th slot result, and $z = [z_1, z_2, ..., z_M]$ for all slots output sequence. The MAP criterion gives the best estimate for the transmitted symbol using map

$$h(z) = \arg\max_i \left[p_{i,z}\right]. \quad (9)$$

Where $p_{i,z}$ is the joint probability that the $i$-th signal is transmitted and output sequence is $z$. It leads to the probability of correct detection

$$P_c = \sum_z \max_i p_{i,z}. \quad (10)$$

### B. Strategy Optimization

In order to determine the control parameters $\beta_l$ in each time slot, we rewrite the probability of correct decision to define the reward-to-go function [16]

$$J_k(s_k(z'), \beta_{k+1}) = \sum_{z'' \in Z_{M-k}} p_{h([z',z'']),[z',z'']}, \ (k=0,...,M-1). \quad (11)$$

Where $z'$ is the photon counting output before the $k+1$-th measurement. For the case of $k = 1$, $z'$ values can be either 0 or 1, which means the first time slot output 0 and 1 respectively. Generally the range of $z'$ is $F^{\otimes k}$, where $F$ is binary field {0, 1}. $z''$ is the next $M - k$ slots output, and $Z_{M-k} = F^{\otimes(M-k)}$. $s_k = [p_{1,z'}, p_{2,z'}, ..., p_{M,z'}]$ is the system state and each element is the joint probability $p_{i,z'}$. $\beta_k \in \Sigma$ is the control parameter, where

$$\Sigma = \begin{cases} \{0, -\alpha\}, & \text{for OOK} \\ \{-\alpha, \alpha\}, & \text{for BPSK.} \end{cases} \quad (12)$$

At first, let us define some symbols. We associate each leaf of strategy tree, shown in Fig. 1 (*b*), with a probability $p_{h(z),z}$ and each $z'$ corresponds to a tree node $n$. For each node $n$, we can construct a sub-tree $T$ whose root node is $n$. Define the probability sum of $T$ as the sum of all its leaves. Then the meaning of $J_k$ for the $z'$ is the probability of the sub-tree $T$.

For convenient we define

$$J_M(s_M(z), \beta_{M+1}) = p_{h(z),z}. \quad (13)$$

It is easy to verify that

$$J_0(s_0, \beta_1) = P_c, \quad (14)$$

and

$$\begin{aligned} J_k(s_k(z'), \beta_{k+1}) &= \sum_{z'' \in Z_{M-k}} p_{h([z',z'']),[z',z'']} \\ &= \sum_{z'' \in Z_{M-k-1}} p_{h([z',0,z'']),[z',0,z'']} + \sum_{z'' \in Z_{M-k-1}} p_{h([z',1,z'']),[z',1,z'']} \\ &= J_{k+1}(s_{k+1}([z',0]), \beta^0_{k+2}) + J_{k+1}(s_{k+1}([z',1]), \beta^1_{k+2}) \end{aligned} \quad (15)$$

The states

$$\begin{aligned} s_{k+1}([z',0]) &= \left[p_{1,[z',0]}, ..., p_{M,[z',0]}\right] \\ &= \left[p_{0|c^1_{k+1}} p_{1,z'}, ..., p_{0|c^M_{k+1}} p_{M,z'}\right], \\ s_{k+1}([z',1]) &= \left[p_{1,[z',1]}, ..., p_{M,[z',1]}\right] \\ &= \left[p_{1|c^1_{k+1}} p_{1,z'}, ..., p_{1|c^M_{k+1}} p_{M,z'}\right]. \end{aligned} \quad (16)$$

Where $c^i_k$ stands for the $i$-th signal symbol in the $k$-th time slot. For MPPM and OOK quantum coding signals, the value 0 stands for no pulse and 1 for pulse; For BPSK quantum coding signals, the value 0 stands for $|-\alpha\rangle$ and 1 for $|\alpha\rangle$.

In order to make the correct probability maximum, we just need to maximize $J_0$ over $\beta_1$. According to recursive relationship (15), we need to solve two sub-problems, namely

$$J^*_k(s(z'),\beta_{k+1}) = \max_{\beta_{k+2}} J^*_{k+1}(s_{k+1}([z',0]),\beta_{k+2}) \\ + \max_{\beta_{k+2}} J^*_{k+1}(s_{k+1}([z',1]),\beta_{k+2})$$ (17)

Where $J^*_k(s(z'),\beta_{k+1})$ is the reward-to-go function $J_k(s(z'),\beta_{k+1})$ with the best control parameters sequence $[\beta^*_{k+2}, \beta^*_{k+3}, \ldots, \beta^*_M]$. And the best control parameters are given by

$$\beta^*_{k+1} = \mathrm{argmax}_{\beta_{k+1}} J^*_k(s_k([z',i]),\beta_{k+1}).$$ (18)

In summary, the optimization algorithm can be described by the following procedure:
1. Set states $s_0 = [1/M,\ldots,1/M]$.
2. For $\beta_1$ in control parameter set $\Sigma$, calculate $J_0(s_0, \beta_1)$ and select the value to maximize $J_0$.

And for each $k=0,1,\ldots,M$, $J_k(s_k, \beta_{k+1})$ and the best control parameters can be optimized by following recursion procedure:
1. If $k = M$, return maximum value in $s_k$, which means using MAP criterion to select the hypothesis for this output sequence z by (9).
2. For other situations, calculate the new states $s_{k+1}([z',0])$ and $s_{k+1}([z',1])$ for the given control parameter $\beta_{k+1}$ as in (14) respectively, and the conditional probabilities in (14) can be calculated by (5) – (8).
3. Calculate $J_{k+1}(s_{k+1}([z', 0]), \beta_{k+2})$ and $J_{k+1}(s_{k+1}([z', 1]), \beta_{k+2})$ respectively for each $\beta_{k+2}$ in set $\Sigma$, and return the value by (17). And the best control parameter for each sub-problem is given by (18).

### C. Numerical Simulation Results

We ran the optimization algorithm and numerical simulation for 2-4-PPM, (7, 4) hamming code OOK and BPSK quantum coding signals, see Fig. 2. Our CPN receivers are compared with the traditional receiver and square root measurement (SRM). The SRM is a nearly optimal measurement.

For 2-4-PPM and (7, 4) Hamming code OOK signals, the performance of our CPN receiver outperforms the DD receiver for any average photon number. For (7, 4) Hamming code BPSK signals, our CPN receiver is compared with homodyne detection (HD) receiver, and it outperforms the homodyne detection receiver when the light is bright.

The reason why CPN works is that the DD receivers directly detect each pulse independently, and use maximum-likelihood criterion to decode the message sent by transmitter, while the CPN receivers do these two tasks together. So it can use previous knowledge to change the detection method in the next time slot, which is known as joint detection.

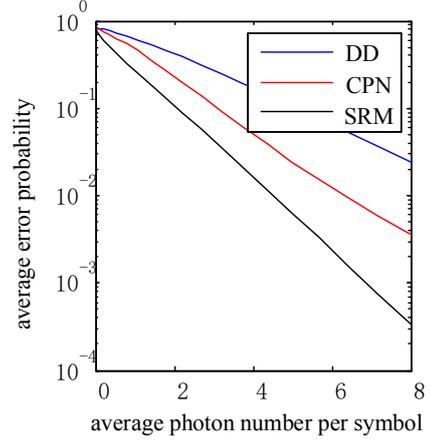

(a)

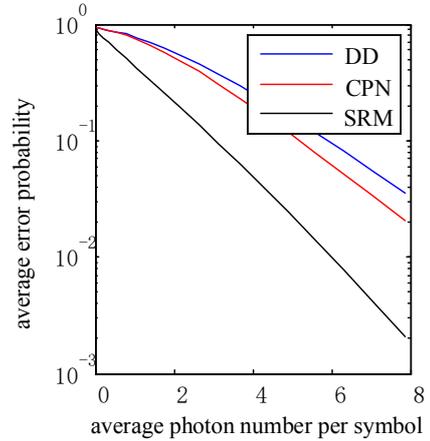

(b)

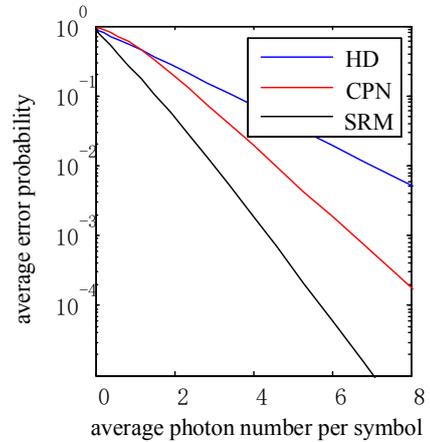

(c)

Fig. 2. Performances comparison of different receivers for different signals. (a) 2-4-PPM signals. The curves, from top to bottom, correspond to direct detection (DD) and decoding with maximum-likelihood criterion (blue), conditional pulse nulling (CPN) receiver (red), and square root measure (SRM) (black). (b) (7, 4) Hamming code OOK signals. Each line is DD and decoding with maximum-likelihood criterion (blue), CPN (red) and SRM (black). (c) (7, 4) Hamming code BPSK signals. The curves are homodyne detection (HD) and decoding with maximum-likelihood criterion (blue), CPN (red) and SRM (black) respectively.

IV. CONCLUSIONS

In this paper, the conditional nulling receivers for multi-pulse PPM and binary quantum coding signals are proposed. And an algorithm is applied to optimize the control strategy. By using numerical simulation, we proved that the CPN receivers can further reduce the average error probability below the traditional detection method for both MPPM and binary quantum coding signals. So they are potential to be applied to the deep space communication system and other optical communication systems to improve the band-utilization efficiency and the channel capacity.

On the other hand, the control parameters are restricted within only two value in our case. If we free this constrain, how to optimize the best control strategy efficiently is a subject for further work. Besides, if we use Dolinar receiver in each time slot, the result for CPN receiver is also worthy of exploring.